\documentclass[aps,prd,amsmath,amssymb,superscriptaddress,preprintnumbers,preprint]{revtex4-1}
\usepackage[utf8]{inputenc} 
\usepackage{graphicx}
\usepackage{dcolumn}
\usepackage{bm}
\usepackage[bookmarks, pagebackref=false]{hyperref}
\usepackage[usenames,dvipsnames]{xcolor}
\usepackage[normalem]{ulem}
\definecolor{rossoCP3}{cmyk}{0,.88,.77,.40}
\definecolor{blaa}{rgb}{0.2,0.2,0.6}
\hypersetup{colorlinks, 
	bookmarksopen, 
	bookmarksnumbered,
	citecolor=blaa, 		
	linkcolor=rossoCP3,	
	urlcolor=rossoCP3,			
}
\usepackage{multirow}

\usepackage{natbib}
\usepackage{slashed}
\newcolumntype{x}[1]{>{\centering\arraybackslash\hspace{0pt}}p{#1}}

\graphicspath{{./figs/}}

\begin{document}
 
\title{ \LARGE  \color{rossoCP3} The W boson mass weighs in\\ on the non-standard Higgs}
\author{Giacomo {\sc Cacciapaglia}}
\thanks{{\scriptsize Email}: \href{mailto:g.cacciapaglia@ipnl.in2p3.fr}{g.cacciapaglia@ipnl.in2p3.fr}; {\scriptsize ORCID}: \href{https://orcid.org/0000-0002-3426-1618}{0000-0002-3426-1618}}
\affiliation{Institut de Physique des 2 Infinis de Lyon (IP2I), UMR5822, CNRS/IN2P3,  F-69622 Villeurbanne Cedex, France}
\affiliation{University of Lyon, Universit\'e Claude Bernard Lyon 1, F-69001 Lyon, France}
 
\author{Francesco {\sc Sannino}}
\thanks{{\scriptsize Email}: \href{mailto:sannino@cp3.sdu.dk}{sannino@cp3.sdu.dk}; {\scriptsize ORCID}: \href{https://orcid.org/0000-0003-2361-5326}{ 0000-0003-2361-5326}}
\affiliation{CERN, Theoretical Physics Department, 1211 Geneva 23, Switzerland}
\affiliation{Scuola Superiore Meridionale, Largo S. Marcellino, 10, 80138 Napoli NA, Italy}
\affiliation{Dipartimento di Fisica, E. Pancini, Universit\'a di Napoli Federico II, INFN-Napoli, Complesso Universitario di Monte S. Angelo Edificio 6, via Cintia, 80126 Napoli, Italy}
\affiliation{CP$^3$-Origins and Danish-IAS, Univ. of Southern Denmark, Campusvej 55, 5230 Odense M, Denmark}

\begin{abstract}  
We consider the implications of the CDF collaboration high-precision measurement of the W boson mass on models with a non-standard Higgs. We show that this requires an enhancement of 3-10\% in the non-standard Higgs coupling to the gauge bosons. This is naturally accommodated in dynamical models such as the dilaton Higgs, the Technicolor and glueball Higgs. The needed composite scale between 2 and 3 TeV can also explain the muon g-2 anomaly, as well as possible violations of lepton flavour universality. 
\end{abstract}
 
\maketitle

The CDF collaboration measured the $W$ boson mass $M_W$ using data relative to 8.8 inverse femtobarns (fb$^{-1}$) of integrated luminosity, collected in proton-antiproton collisions at an energy in the center-of-mass of $1.96$~TeV, via the CDF II detector at the Fermilab Tevatron collider. With a sample of about 4-million $W$ bosons, they obtained \cite{
doi:10.1126/science.abk1781}
\begin{equation} \label{eq:CDF}
    \left. M_W \right|_{\rm CDF} = 80,433.5 \pm 6.4_{\rm stat} \pm 6.9_{\rm syst}= 80,433.5 \pm 9.4\, {\rm MeV}  \ . 
\end{equation}
There are two striking results associated to this new measurement. 

The first is that the central value is larger than expected, leading to a strong tension with the Standard Model (SM) expectation \cite{ParticleDataGroup:2020ssz}:
\begin{equation} \label{eq:SM}
    \left. M_W \right|_{\rm SM} = 80,357 \pm 4_{\rm inputs} \pm 4_{\rm theory}  \, {\rm MeV}  \ . 
\end{equation}
The SM result derives from symmetries (mainly the custodial symmetry of the Higgs sector) and a set of high-precision measurements that include the Higgs and Z boson masses, the top-quark mass, the electromagnetic coupling, the muon lifetime and collider asymmetries, which serve as input to the analytic computations. The estimate of the SM expected value of $M_W$ is affected by uncertainties in the input data and by missing higher-order perturbative computations (theory).  All in all, the tension between the CDF measurement and the SM can be quantified to 7 standard deviations (7$\sigma$) \cite{doi:10.1126/science.abk1781}.

The second striking result is that the accuracy of the new CDF measurement exceeds that of all previous measurements combined, coming from the Large Electron Positron collider (LEP)  and previous Tevatron analyses \cite{ALEPH:2010aa}:
\begin{equation}
    \left. M_W \right|_{\rm LEP} = 80,385 \pm 15 \, {\rm MeV}  \ .
    \end{equation}
More recently, the ATLAS collaboration at the Large Hadron Collider (LHC) presented a new result with competitive accuracy \cite{ATLAS:2017rzl}: \begin{equation}
   \left. M_W \right|_{\rm ATLAS} = 80,370 \pm 19 \, {\rm MeV}  \ . 
    \end{equation}
A simple average of these 3 measurements yields: 
\begin{equation} \label{eq:AVG}
    \left. M_W \right|_{\rm AVG} = 80,396\pm 9 \, {\rm MeV} \ .
\end{equation}
Comparing our naive average with the SM in Eq.~\eqref{eq:SM}, the discrepancy is reduced to around 4 standard deviations.

\vspace{0.5cm}

\begin{figure}
    \centering
\includegraphics[width=0.9\textwidth]{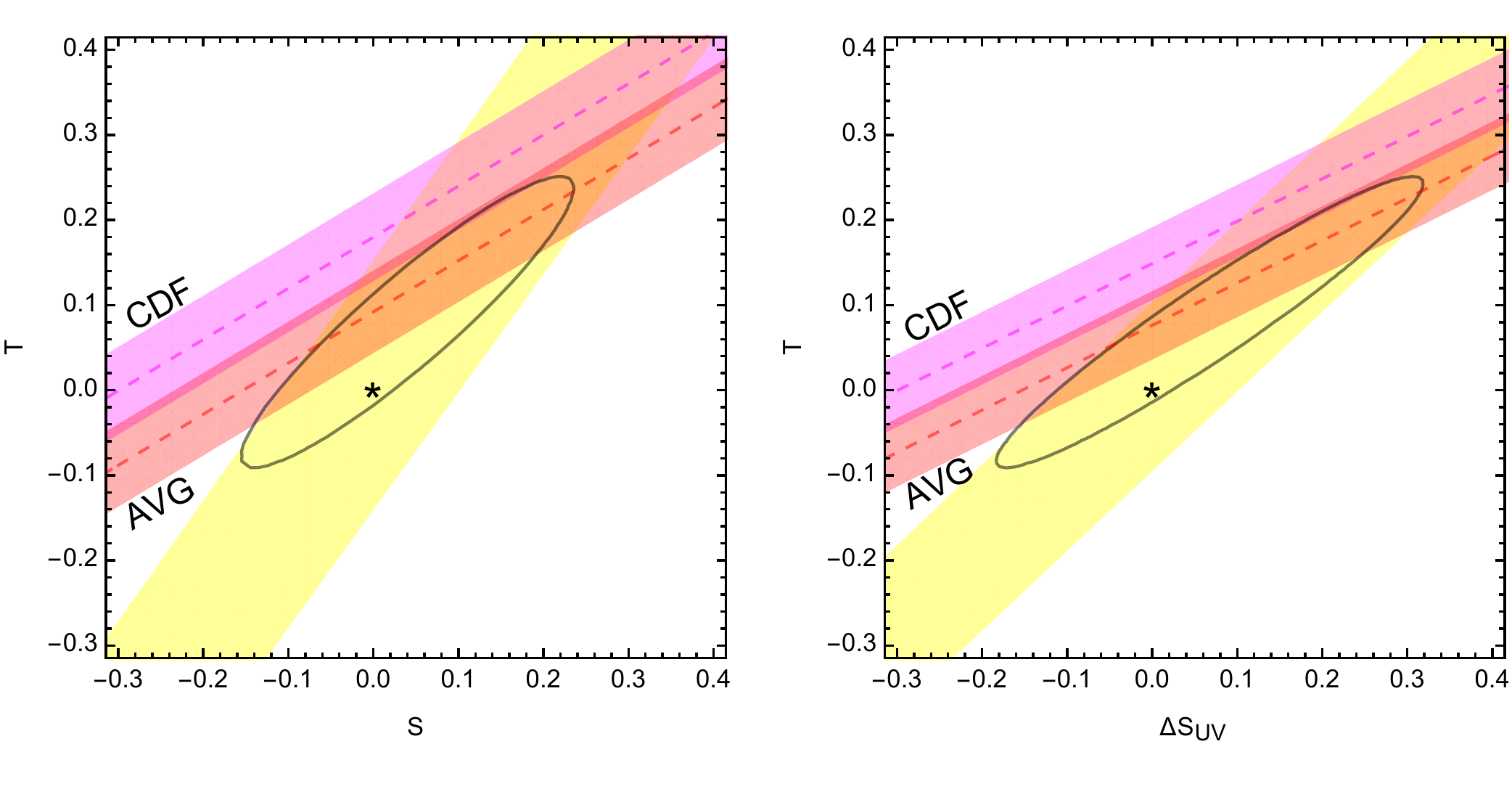}
    \caption{Comparison of the new CDF measurement of the $W$ mass (magenta band) to the precision measurements, represented by the ellipse. The yellow band indicates the bound stemming from asymmetries and direct $s_W^2$ determination. The red band represents our naive average that includes the CDF result. All bounds are shown at 95\% confidence level (i.e. 2$\sigma$).}
    \label{fig:ST}
\end{figure}

Taking these measurements at face value, the deviation from the SM can be accounted for by new physics beyond the SM. In particular, the $W$ mass is very sensitive to contributions to the vacuum polarisations of the electroweak bosons, encoded in the oblique parameters. This correction can be expressed in terms of the Altarelli-Barbieri epsilon parameters \cite{Altarelli:1990zd}:
\begin{equation}
    \frac{M_W^2}{M_Z^2} = \left. \frac{M_W^2}{M_Z^2} \right|_{\rm Born} \ast  (1+1.34\ \epsilon_1 - 0.86\ \epsilon_3)\,, 
\end{equation}
or via the oblique $S$ and $T$ parameters \cite{Peskin:1990zt,Peskin:1991sw}. The latter set only includes the contribution of new physics, and can therefore be defined only once the SM part is  known. After the discovery of the Higgs boson \cite{ATLAS:2012yve,CMS:2012qbp} and the precise measurement of its mass \cite{ATLAS:2015yey}, the new physics contributions to the oblique parameters can be clearly defined:
\begin{equation}
    \delta \epsilon_1 = \alpha T\,, \qquad \delta \epsilon_3 = \alpha S\,,
\end{equation}
where $\alpha$ is the electromagnetic coupling constant at the $Z$ mass. The correction to the $W$ mass stemming from new physics can therefore be expressed in terms of  $S$ and $T$ via the following numerical approximate formula \cite{Altarelli:1994iz}:
\begin{equation}
    \Delta M_W \approx 300\ \text{MeV} \ast (1.43 \ T - 0.86\ S)\,,
\end{equation}
which we will use in our numerical analysis.
We now compare this correction to the deviation provided by the new CDF measurement and by our naive average:
\begin{eqnarray}
    \left. \Delta M_W \right|_{\rm CDF} & =\left. M_W \right|_{\rm CDF} - \left. M_W \right|_{\rm SM}= 76  \pm 11\, \text{MeV}\,,  \nonumber \\   \left. \Delta M_W \right|_{\rm AVG} & = \left. M_W \right|_{\rm AVG} - \left. M_W \right|_{\rm SM}= 39 \pm 10\, \text{MeV}\,.
\end{eqnarray}
This is illustrated by the magenta (CDF) and red (AVG) bands at 2$\sigma$ in the left panel of Fig.~\ref{fig:ST}, compared to the bound coming from the asymmetries and the direct determination of $s_W^2$ in the SM \cite{Haller:2018nnx}, represented by the yellow band. The ellipse stems from the fit to the oblique corrections without including the CDF measurement \footnote{$S = 0.04 \pm 0.08$ and $T = 0.08 \pm 0.08$ with a correlation of $0.91$ \cite{Haller:2018nnx}.}.

For models featuring a non-standard Higgs bosons, it is the coupling of the Higgs to gauge bosons that is mainly responsible for corrections to the oblique parameters as well as the value of the $W$ mass. This is efficiently encoded in the parameter $\kappa_V$, which has the same value for $W$ and $Z$ assuming an underlying custodial model. This parameter can be expressed at tree level in terms of observables as follows: 
\begin{equation}
    \kappa_V^2 = \frac{\sigma_{\rm VBF}}{\sigma_{\rm VBF}^{\rm SM}} \equiv \frac{\Gamma_{h\to WW/ZZ}}{\Gamma_{h\to WW/ZZ}^{\rm SM}}\,,
\end{equation}
hence the SM corresponds to $\kappa_V = 1$. 
The well-know relation between $\kappa_V$ and the oblique parameters is:
\begin{equation} \label{eq:SandT}
T = - \frac{3}{16 \pi c_W^2} (1-\kappa_V^2)  \ln \frac{\Lambda^2}{m_h^2}\,, \qquad 
S = \frac{1}{12 \pi} (1-\kappa_V^2)  \ln \frac{\Lambda^2}{m_h^2} + \Delta S_{\rm UV}\,,
\end{equation}
where we have included an unknown contribution to the $S$ parameter stemming from UV physics. Assuming the new UV physics to be custodial, it does not contribute to $T$.
One can express $S$ in terms of $T$ as follows:
\begin{equation}
S = - \frac{4}{9} c_W^2  T + \Delta S_{\rm UV}\,,
\end{equation}
and plot the general bounds in terms of $T$ (coming from the Higgs coupling modifications) and the UV contribution to $S$. This model-independent result is shown in the right panel of Fig.~\ref{fig:ST}. The take-home message is that the independent measurement of the CDF W mass (magenta band) and $s^2_W$ (yellow band) prefer a small and positive $\Delta S_{UV}$ for positive $T > 0.1$. 

\begin{figure}
    \centering
\includegraphics[width=0.9\textwidth]{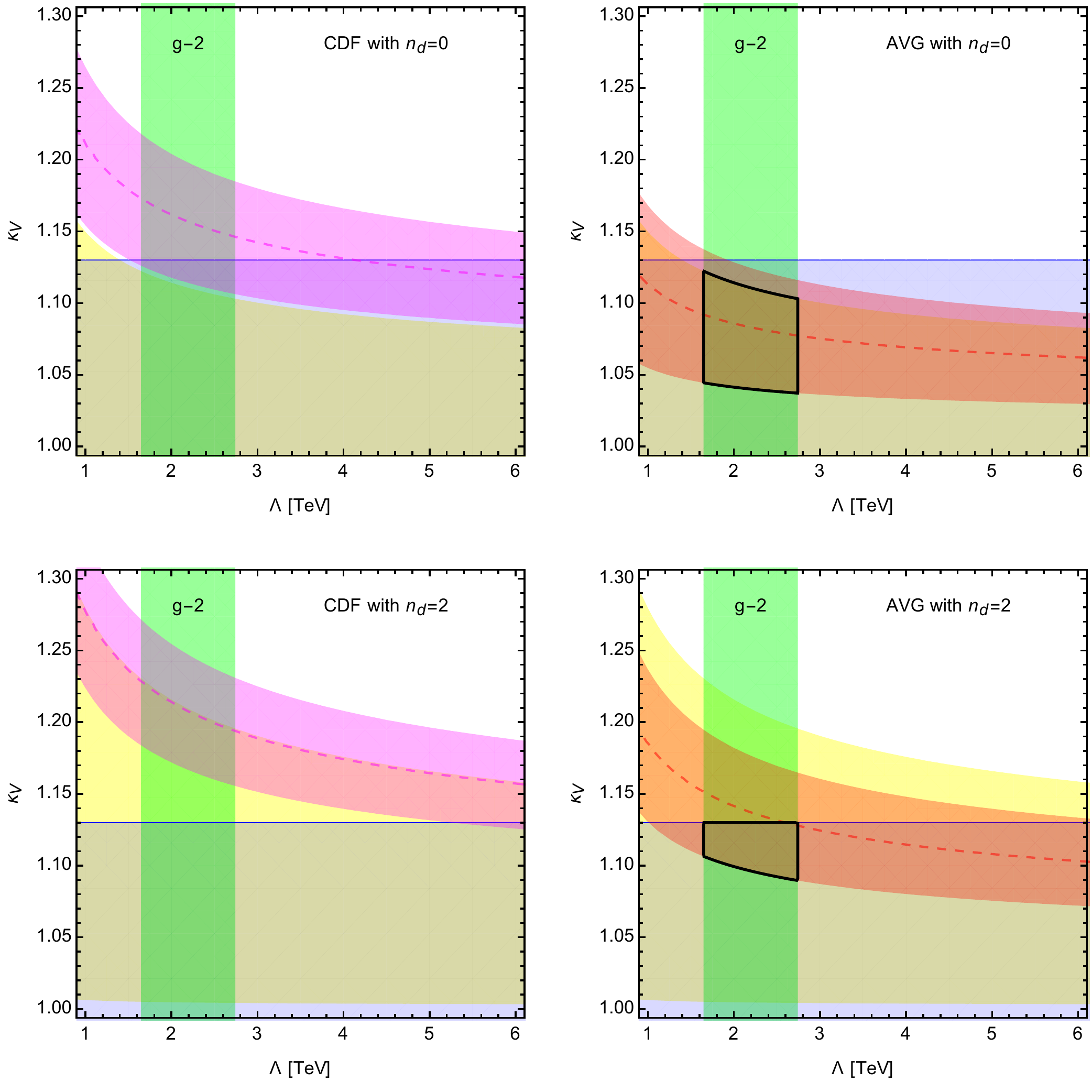}
    \caption{Bounds at 95\% Confidence Level (2$\sigma$) from electroweak precision (asymmetries and $s_W^2$ in yellow) and the direct $\kappa_V$ measurement (blue) compared to the $M_W$ measurements from CDF (magenta in the left panels) and the naive average (red in the right panels). The green vertical band corresponds to the region preferred by the muon $g-2$ anomaly. The top panels correspond to vanishing UV contribution to $S$ ($n_d = 0$), while the bottom ones assume $n_d = 2$ ($\Delta S_{\rm UV} = 1/3\pi$). The regions compatible with all bounds are highlighted by the black contour.}
    \label{fig:models}
\end{figure}

We can now express the same bounds in terms of the non-standard Higgs coupling $\kappa_V$ and the scale of new physics $\Lambda$, as shown in Fig.~\ref{fig:models}. The UV contribution to $S$ has been parameterised as
\begin{equation}
\Delta S_{\rm UV} = n_d \frac{1}{6\pi}\,,
\end{equation}
which roughly counts the number of SU(2) weak doublets $n_d$ present in the UV theory \cite{Peskin:1991sw}.
In Fig.~\ref{fig:models} we also include the limit on $\kappa_V$ stemming from direct measurements, where the most constraining value is due to the ATLAS collaboration with a combination of the full Run-I data and various integrated luminosities (ranging from 24.5 to 79.8 fb$^{-1}$) for Run-II data \cite{ATLAS:2019nkf}:
\begin{equation}
    \kappa_V = 1.05 \pm 0.04\,,
\end{equation}
which is also compatible with the CMS results \cite{CMS:2018uag}.
What we learn from the plots on the left column is that the CDF measurement alone is incompatible with other bounds as it would need a too large $\kappa_V$ in order to generate a large enough $T$ compatible with the other precision measurements. Only regions with large new physics scale could be marginally compatible if a positive $\Delta S_{\rm UV}$ is present.
The naive average, instead, is fully compatible and leads to $\kappa_V \gtrsim 1.03$. Larger values of $\kappa_V$ are needed for increasing $\Delta S_{\rm UV}$ (as shown in the bottom-right panel), so that a rough  upper limit of $n_d \lesssim 4$ can be obtained.

Interestingly, the anomaly in the muon $g-2$ \cite{Muong-2:2021ojo} can be  related to a new physics contribution appearing at the scale $\Lambda$ via the naive relation
\begin{equation}
    \Delta a_\mu = \frac{m_\mu^2}{\Lambda^2}\,.
\end{equation}
This relation is the right type of estimate for strongly coupled models \cite{Cacciapaglia:2021gff}, where no loop factors arise and $\Lambda$ is directly associated to the scale where new resonances could appear, like a techni-rho.  
The vertical green band in Fig.~\ref{fig:models} indicates the 2$\sigma$ preferred region, where we do not include the recent BMW lattice \cite{Borsanyi:2020mff} determination in the SM world average \cite{Aoyama:2020ynm}. In general, this anomaly is compatible with the new average for the $W$ mass as long as $\Lambda$ is between $1.6$ to $2.8$~TeV. If we consider a reduced anomaly, as driven by the lattice results, the scale is pushed to slightly higher values \cite{Cacciapaglia:2021gff}, without spoiling the compatibility with the $W$ mass measurement. Following \cite{Cacciapaglia:2021gff} one can also address the observed lepton flavour non-universal anomalies \cite{LHCb:2021trn} via fundamental partial composite models \cite{Sannino:2016sfx}, further reviewed in \cite{Cacciapaglia:2020kgq}, with different muon left and right couplings. 


\vspace{0.5cm}


The Higgs boson with non-standard interactions, which we analysed so far, arises in many new physics realisations. In this letter we are interested in dynamical models, where a Higgs-like boson arises as a composite state of a more fundamental interaction. The low energy properties can be described in terms of an effective field theory, which is controllable thanks to some symmetries of the underlying theory. In the following, we will consider four classes of model: a pseudo-dilaton \cite{Yamawaki:1985zg}, a dynamical glueball \cite{Sannino:2015yxa} or composite Higgs \cite{Hong:2004td}, and Goldstone/Holographic composite Higgs \cite{Kaplan:1983fs,Contino:2003ve}. The non-standard interactions are captured by the following master Lagrangian \cite{Sannino:2015yxa}:
\begin{eqnarray}
{\cal L} &=& {\cal L}_{\overline{\rm SM}}
+\xi_V \left(1+2 \kappa_V \frac{h}{v}+\kappa_{2V} \frac{h^2}{v^2}\right)\frac{v^2}{4}{\rm Tr}\ D_\mu U^\dagger D^\mu U 
+ \frac{1}{2} \partial_{\mu} h \partial^{\mu} h 
 -  \frac{\tilde{m}^2_h}{2} h^2 \left( 1 +  V_{0,1} \lambda_{3h} \frac{h}{v}\right) \
 \nonumber \\
&& \phantom{xxxxx} -
\ \frac{\tilde{y}_t v}{\sqrt{2}} \xi_t \left(1+\kappa_t \frac{h}{v}\right)
\Bigg[\overline{q}_L\ U\ \Bigg(\frac{1}{2}+T^3\Bigg)\ q_R + {\rm h.c.} \Bigg] \nonumber \\
&&\phantom{xxxxxxxxxxxxxxxxxxxx} - \frac{\tilde{y}_b v}{\sqrt{2}} \xi_b \left(1+\kappa_b \frac{h}{v}\right)
\Bigg[\overline{q}_L\ U\ \Bigg(\frac{1}{2}-T^3\Bigg) \ q_R + {\rm h.c.} \Bigg] + \cdots 
\label{eq:Lmodels}
\end{eqnarray}
where $U = \text{exp} \left( i 2 \pi^a T^a/v\right)$ is the usual non-linear map of the Goldstones $\pi^a$ produced by the breaking of the electroweak symmetry.

The difference among the four classes lays in different counting schemes for the couplings in the above Lagrangian. A useful tool consists in relying on a large-$N$ counting, where $N$ is the number of colours in the confining underlying theory \cite{Sannino:2015yxa}. In the glueball case, the Higgs can be envisioned as the lightest glueball state of a new Yang-Mills theory with a new $N$-dependent string tension proportional to $\Lambda_H$, scale not automatically related to the electroweak scale $v=246$~GeV. A composite Higgs \cite{Hong:2004td} in Technicolor-like theories \cite{Weinberg:1975gm,Susskind:1978ms}, instead, is associated to a fermion-antifermion bound meson. The counting now depends on the fermion representation \cite{Sannino:2009za}. Here we express the counting with respect to a reference theory based on SU($\overline{N}$) gauge with fermions in the fundamental.
In the dilaton case, the scalar state is associated to a spontaneoulsy broken conformal dynamics, and its couplings are associated to a single scale $F \equiv N \Lambda_H$. Finally, in the case of the Goldstone/Holographic Higgs \cite{Contino:2003ve}, the counting is based on the misalignment \cite{Kaplan:1983fs} between the electroweak scale and the compositeness scale $f$, which can be expressed in terms of an angle \cite{Dugan:1984hq}, $v = f s_\theta$  with $s_\theta \ll 1$.

\begin{table}[htb]
\begin{tabular}{l||c|c|c|c|c|c|c|}
 & $\xi_V$ & $\kappa_V$ & $\kappa_{2V}$ & $\lambda_{3h}$ & $\xi_f$ & $\kappa_f$ & higher orders \\
 \hline\hline
Glueball \phantom{\Big(} & $1$ & $\displaystyle \frac{r_\pi v}{N \Lambda_H}$ & $\displaystyle \frac{s_\pi v^2}{N^2 \Lambda_H^2}$ & $\displaystyle \frac{v}{N \Lambda_H}$ & $1$ & $\displaystyle \frac{r_f v}{N \Lambda_H}$ & $\displaystyle \frac{1}{4\pi v},\, \frac{\partial^2}{\Lambda_H^2}$ \\ \hline
Technicolor-like \phantom{\Big(} & $\displaystyle \frac{N}{\overline{N}}$ & $\displaystyle r_\pi \sqrt{\frac{\overline{N}}{N}}$ & $\displaystyle s_\pi \frac{\overline{N}}{N}$ & $\sqrt{\frac{\overline{N}}{N}}$ & $\sqrt{\frac{N}{\overline{N}}}$ & $\displaystyle r_f \sqrt{\frac{\overline{N}}{N}}$ & $\displaystyle \frac{1}{4\pi v},\, \frac{\partial^2}{v^2}$ \\ \hline
pseudo-dilaton \phantom{\Big(} & $1$ & $\displaystyle \frac{v}{N \Lambda_H}$ & $\displaystyle \frac{v^2}{N^2 \Lambda_H^2}$ & $\displaystyle \frac{v}{N \Lambda_H}$ & $1$ & $\displaystyle \frac{v}{N \Lambda_H}$ &  \\ \hline
Goldstone/Holographic & $1$ & $c_\theta$ & $c_{2\theta}$ & $c_\theta$ & $1$ & $c_\theta$ & $\displaystyle \frac{1}{4 \pi f},\, \frac{\partial^2}{f^2}$ \\ \hline
\end{tabular}
\caption{Couplings in Eq.~\eqref{eq:Lmodels} in the four classes of models \cite{Sannino:2015yxa}.}
\label{tab:1}
\end{table}

The relevant couplings in the four classes are summarised in Table~\ref{tab:1}, where the Goldstone/Holographic couplings correspond to the minimal model of Ref.~\cite{Agashe:2004rs}. Note that in the Goldstone/Holographic case, the couplings to gauge bosons are always reduced with respect to the SM values, as $\kappa_V = c_\theta \leq 1$. This is a universal property of Goldstone Higgs models \cite{Liu:2018vel}. Only couplings to fermions can be enhanced compared to the SM values, for specific choices of the model details \cite{Hashimoto:2017jvc}.
In all the other cases, $\kappa_V$ is naturally larger than unity when the new scale is somewhat smaller that $v$, or the coupling $r_\pi > 1$. As the new $W$ mass measurement points towards a positive deviation of  $\kappa_V$ from unity of  circa $ 3 \div 10$\%, this leads to a composite scale  $\Lambda \sim 4 \pi N \Lambda_H \approx 4 \pi v$, which is in agreement with the $g-2$ preferred region.    Note that the coupling to fermions is also expected to receive a similar enhancement, that is  compatible with the current experimental measurement $\kappa_f = 1.05 \pm 0.09$ \cite{ATLAS:2019nkf}. The dilaton case is most predictive, as other measurable couplings can be determined once the scale of new physics is fixed, for instance the Higgs trilinear coupling that will be measured at future colliders.


\vspace{0.5cm}

To summarise, in this letter we discussed the impact on non-standard Higgs models of the new $W$ mass measurement from the CDF collaboration, which has not only the highest precision to date, but also a central value significantly higher than previous measurements and than the SM fit result.
Our analysis includes composite models, but it is not limited to them. Firstly, we noted that the CDF measurement on its own, which has a 7$\sigma$ discrepancy to the SM fit, is very hard to accommodate as it would require a large departure in the Higgs couplings to $W$ and $Z$ bosons, unless large custodial violations are present in the UV model. However, a naive average with the LEP, previous Tevatron and ATLAS $W$ mass measurements, while reducing the anomaly to 4$\sigma$, renders the overall result easier to accommodate. 

The main implication stemming from the new average including the CDF result is that the coupling of the non-standard Higgs to $W$ and $Z$ bosons is required to be larger than the SM value by 3--10\%. This is rather hard to accommodate in weakly coupled models and in models of Goldstone/Holographic Higgs, where a reduction of this coupling is usually obtained, unless a sizeable source of custodial violation is present in the UV theory giving a positive contribution to $T$. However, this feature can be naturally achieved in composite models where the Higgs emerges as a resonance, as in Technicolor-like theories or pseudo-dilaton models. Furthermore, a similar enhancement is expected in the fermion couplings of the Higgs. The High-Luminosity LHC and future Higgs factories will be able to establish the non-standard nature of the Higgs couplings, as they are expected to reach a precision of about 2\% \cite{Cepeda:2019klc} and $<$1\% \cite{FCC:2018byv}, respectively. Furthermore, the anomaly in the muon $g-2$ measurement is also compatible with the dynamical origin of the Higgs, if the new composite scale lies between $1.6$ to $2.8$~TeV, where new resonances like a spin-1 techni-rho are expected.

In conclusion, it is tantalising that the anomalies that are emerging in the precision measurement of the SM properties  conspire towards a coherent picture, where the Higgs boson emerges from a dynamical composite sector at a few TeV scale. This scenario will be discovered or excluded by the end of the LHC program, and most certainly by the next high-precision colliders, whose construction is being discussed.

\bibliography{g2biblio}

\begin{thebibliography}{36}%
\makeatletter
\providecommand \@ifxundefined [1]{%
 \@ifx{#1\undefined}
}%
\providecommand \@ifnum [1]{%
 \ifnum #1\expandafter \@firstoftwo
 \else \expandafter \@secondoftwo
 \fi
}%
\providecommand \@ifx [1]{%
 \ifx #1\expandafter \@firstoftwo
 \else \expandafter \@secondoftwo
 \fi
}%
\providecommand \natexlab [1]{#1}%
\providecommand \enquote  [1]{``#1''}%
\providecommand \bibnamefont  [1]{#1}%
\providecommand \bibfnamefont [1]{#1}%
\providecommand \citenamefont [1]{#1}%
\providecommand \href@noop [0]{\@secondoftwo}%
\providecommand \href [0]{\begingroup \@sanitize@url \@href}%
\providecommand \@href[1]{\@@startlink{#1}\@@href}%
\providecommand \@@href[1]{\endgroup#1\@@endlink}%
\providecommand \@sanitize@url [0]{\catcode `\\12\catcode `\$12\catcode
  `\&12\catcode `\#12\catcode `\^12\catcode `\_12\catcode `\%12\relax}%
\providecommand \@@startlink[1]{}%
\providecommand \@@endlink[0]{}%
\providecommand \url  [0]{\begingroup\@sanitize@url \@url }%
\providecommand \@url [1]{\endgroup\@href {#1}{\urlprefix }}%
\providecommand \urlprefix  [0]{URL }%
\providecommand \Eprint [0]{\href }%
\providecommand \doibase [0]{http://dx.doi.org/}%
\providecommand \selectlanguage [0]{\@gobble}%
\providecommand \bibinfo  [0]{\@secondoftwo}%
\providecommand \bibfield  [0]{\@secondoftwo}%
\providecommand \translation [1]{[#1]}%
\providecommand \BibitemOpen [0]{}%
\providecommand \bibitemStop [0]{}%
\providecommand \bibitemNoStop [0]{.\EOS\space}%
\providecommand \EOS [0]{\spacefactor3000\relax}%
\providecommand \BibitemShut  [1]{\csname bibitem#1\endcsname}%
\let\auto@bib@innerbib\@empty
\bibitem [{\citenamefont {Aaltonen}\ \emph {et~al.}(2022)\citenamefont
  {Aaltonen} \emph {et~al.}}]{doi:10.1126/science.abk1781}%
  \BibitemOpen
  \bibfield  {author} {\bibinfo {author} {\bibfnamefont {T.}~\bibnamefont
  {Aaltonen}} \emph {et~al.} (\bibinfo {collaboration} {CDF}),\ }\href
  {\doibase 10.1126/science.abk1781} {\bibfield  {journal} {\bibinfo  {journal}
  {Science}\ }\textbf {\bibinfo {volume} {376}},\ \bibinfo {pages} {170}
  (\bibinfo {year} {2022})}\BibitemShut {NoStop}%
\bibitem [{\citenamefont {Zyla}\ \emph {et~al.}(2020)\citenamefont {Zyla} \emph
  {et~al.}}]{ParticleDataGroup:2020ssz}%
  \BibitemOpen
  \bibfield  {author} {\bibinfo {author} {\bibfnamefont {P.~A.}\ \bibnamefont
  {Zyla}} \emph {et~al.} (\bibinfo {collaboration} {Particle Data Group}),\
  }\href {\doibase 10.1093/ptep/ptaa104} {\bibfield  {journal} {\bibinfo
  {journal} {PTEP}\ }\textbf {\bibinfo {volume} {2020}},\ \bibinfo {pages}
  {083C01} (\bibinfo {year} {2020})}\BibitemShut {NoStop}%
\bibitem [{\citenamefont {{ALEPH, CDF, D0, DELPHI, L3, OPAL, SLD, LEP
  Electroweak Working Group, Tevatron Electroweak Working Group, SLD
  Electroweak, Heavy Flavour Groups}}(2010)}]{ALEPH:2010aa}%
  \BibitemOpen
  \bibfield  {author} {\bibinfo {author} {\bibnamefont {{ALEPH, CDF, D0,
  DELPHI, L3, OPAL, SLD, LEP Electroweak Working Group, Tevatron Electroweak
  Working Group, SLD Electroweak, Heavy Flavour Groups}}},\ }\href@noop {} {\
  (\bibinfo {year} {2010})},\ \Eprint {http://arxiv.org/abs/1012.2367}
  {arXiv:1012.2367 [hep-ex]} \BibitemShut {NoStop}%
\bibitem [{\citenamefont {Aaboud}\ \emph {et~al.}(2018)\citenamefont {Aaboud}
  \emph {et~al.}}]{ATLAS:2017rzl}%
  \BibitemOpen
  \bibfield  {author} {\bibinfo {author} {\bibfnamefont {M.}~\bibnamefont
  {Aaboud}} \emph {et~al.} (\bibinfo {collaboration} {ATLAS}),\ }\href
  {\doibase 10.1140/epjc/s10052-017-5475-4} {\bibfield  {journal} {\bibinfo
  {journal} {Eur. Phys. J. C}\ }\textbf {\bibinfo {volume} {78}},\ \bibinfo
  {pages} {110} (\bibinfo {year} {2018})},\ \bibinfo {note} {[Erratum:
  Eur.Phys.J.C 78, 898 (2018)]},\ \Eprint {http://arxiv.org/abs/1701.07240}
  {arXiv:1701.07240 [hep-ex]} \BibitemShut {NoStop}%
\bibitem [{\citenamefont {Altarelli}\ and\ \citenamefont
  {Barbieri}(1991)}]{Altarelli:1990zd}%
  \BibitemOpen
  \bibfield  {author} {\bibinfo {author} {\bibfnamefont {G.}~\bibnamefont
  {Altarelli}}\ and\ \bibinfo {author} {\bibfnamefont {R.}~\bibnamefont
  {Barbieri}},\ }\href {\doibase 10.1016/0370-2693(91)91378-9} {\bibfield
  {journal} {\bibinfo  {journal} {Phys. Lett. B}\ }\textbf {\bibinfo {volume}
  {253}},\ \bibinfo {pages} {161} (\bibinfo {year} {1991})}\BibitemShut
  {NoStop}%
\bibitem [{\citenamefont {Peskin}\ and\ \citenamefont
  {Takeuchi}(1990)}]{Peskin:1990zt}%
  \BibitemOpen
  \bibfield  {author} {\bibinfo {author} {\bibfnamefont {M.~E.}\ \bibnamefont
  {Peskin}}\ and\ \bibinfo {author} {\bibfnamefont {T.}~\bibnamefont
  {Takeuchi}},\ }\href {\doibase 10.1103/PhysRevLett.65.964} {\bibfield
  {journal} {\bibinfo  {journal} {Phys. Rev. Lett.}\ }\textbf {\bibinfo
  {volume} {65}},\ \bibinfo {pages} {964} (\bibinfo {year} {1990})}\BibitemShut
  {NoStop}%
\bibitem [{\citenamefont {Peskin}\ and\ \citenamefont
  {Takeuchi}(1992)}]{Peskin:1991sw}%
  \BibitemOpen
  \bibfield  {author} {\bibinfo {author} {\bibfnamefont {M.~E.}\ \bibnamefont
  {Peskin}}\ and\ \bibinfo {author} {\bibfnamefont {T.}~\bibnamefont
  {Takeuchi}},\ }\href {\doibase 10.1103/PhysRevD.46.381} {\bibfield  {journal}
  {\bibinfo  {journal} {Phys. Rev. D}\ }\textbf {\bibinfo {volume} {46}},\
  \bibinfo {pages} {381} (\bibinfo {year} {1992})}\BibitemShut {NoStop}%
\bibitem [{\citenamefont {Aad}\ \emph {et~al.}(2012)\citenamefont {Aad} \emph
  {et~al.}}]{ATLAS:2012yve}%
  \BibitemOpen
  \bibfield  {author} {\bibinfo {author} {\bibfnamefont {G.}~\bibnamefont
  {Aad}} \emph {et~al.} (\bibinfo {collaboration} {ATLAS}),\ }\href {\doibase
  10.1016/j.physletb.2012.08.020} {\bibfield  {journal} {\bibinfo  {journal}
  {Phys. Lett. B}\ }\textbf {\bibinfo {volume} {716}},\ \bibinfo {pages} {1}
  (\bibinfo {year} {2012})},\ \Eprint {http://arxiv.org/abs/1207.7214}
  {arXiv:1207.7214 [hep-ex]} \BibitemShut {NoStop}%
\bibitem [{\citenamefont {Chatrchyan}\ \emph {et~al.}(2012)\citenamefont
  {Chatrchyan} \emph {et~al.}}]{CMS:2012qbp}%
  \BibitemOpen
  \bibfield  {author} {\bibinfo {author} {\bibfnamefont {S.}~\bibnamefont
  {Chatrchyan}} \emph {et~al.} (\bibinfo {collaboration} {CMS}),\ }\href
  {\doibase 10.1016/j.physletb.2012.08.021} {\bibfield  {journal} {\bibinfo
  {journal} {Phys. Lett. B}\ }\textbf {\bibinfo {volume} {716}},\ \bibinfo
  {pages} {30} (\bibinfo {year} {2012})},\ \Eprint
  {http://arxiv.org/abs/1207.7235} {arXiv:1207.7235 [hep-ex]} \BibitemShut
  {NoStop}%
\bibitem [{\citenamefont {Aad}\ \emph {et~al.}(2015)\citenamefont {Aad} \emph
  {et~al.}}]{ATLAS:2015yey}%
  \BibitemOpen
  \bibfield  {author} {\bibinfo {author} {\bibfnamefont {G.}~\bibnamefont
  {Aad}} \emph {et~al.} (\bibinfo {collaboration} {ATLAS, CMS}),\ }\href
  {\doibase 10.1103/PhysRevLett.114.191803} {\bibfield  {journal} {\bibinfo
  {journal} {Phys. Rev. Lett.}\ }\textbf {\bibinfo {volume} {114}},\ \bibinfo
  {pages} {191803} (\bibinfo {year} {2015})},\ \Eprint
  {http://arxiv.org/abs/1503.07589} {arXiv:1503.07589 [hep-ex]} \BibitemShut
  {NoStop}%
\bibitem [{\citenamefont {Altarelli}\ \emph {et~al.}(1995)\citenamefont
  {Altarelli}, \citenamefont {Barbieri},\ and\ \citenamefont
  {Caravaglios}}]{Altarelli:1994iz}%
  \BibitemOpen
  \bibfield  {author} {\bibinfo {author} {\bibfnamefont {G.}~\bibnamefont
  {Altarelli}}, \bibinfo {author} {\bibfnamefont {R.}~\bibnamefont {Barbieri}},
  \ and\ \bibinfo {author} {\bibfnamefont {F.}~\bibnamefont {Caravaglios}},\
  }\href {\doibase 10.1016/0370-2693(95)00179-O} {\bibfield  {journal}
  {\bibinfo  {journal} {Phys. Lett. B}\ }\textbf {\bibinfo {volume} {349}},\
  \bibinfo {pages} {145} (\bibinfo {year} {1995})}\BibitemShut {NoStop}%
\bibitem [{\citenamefont {Haller}\ \emph {et~al.}(2018)\citenamefont {Haller},
  \citenamefont {Hoecker}, \citenamefont {Kogler}, \citenamefont {M\"onig},
  \citenamefont {Peiffer},\ and\ \citenamefont {Stelzer}}]{Haller:2018nnx}%
  \BibitemOpen
  \bibfield  {author} {\bibinfo {author} {\bibfnamefont {J.}~\bibnamefont
  {Haller}}, \bibinfo {author} {\bibfnamefont {A.}~\bibnamefont {Hoecker}},
  \bibinfo {author} {\bibfnamefont {R.}~\bibnamefont {Kogler}}, \bibinfo
  {author} {\bibfnamefont {K.}~\bibnamefont {M\"onig}}, \bibinfo {author}
  {\bibfnamefont {T.}~\bibnamefont {Peiffer}}, \ and\ \bibinfo {author}
  {\bibfnamefont {J.}~\bibnamefont {Stelzer}},\ }\href {\doibase
  10.1140/epjc/s10052-018-6131-3} {\bibfield  {journal} {\bibinfo  {journal}
  {Eur. Phys. J. C}\ }\textbf {\bibinfo {volume} {78}},\ \bibinfo {pages} {675}
  (\bibinfo {year} {2018})},\ \Eprint {http://arxiv.org/abs/1803.01853}
  {arXiv:1803.01853 [hep-ph]} \BibitemShut {NoStop}%
\bibitem [{Note1()}]{Note1}%
  \BibitemOpen
  \bibinfo {note} {$S = 0.04 \pm 0.08$ and $T = 0.08 \pm 0.08$ with a
  correlation of $0.91$ \cite {Haller:2018nnx}.}\BibitemShut {Stop}%
\bibitem [{\citenamefont {Aad}\ \emph {et~al.}(2020)\citenamefont {Aad} \emph
  {et~al.}}]{ATLAS:2019nkf}%
  \BibitemOpen
  \bibfield  {author} {\bibinfo {author} {\bibfnamefont {G.}~\bibnamefont
  {Aad}} \emph {et~al.} (\bibinfo {collaboration} {ATLAS}),\ }\href {\doibase
  10.1103/PhysRevD.101.012002} {\bibfield  {journal} {\bibinfo  {journal}
  {Phys. Rev. D}\ }\textbf {\bibinfo {volume} {101}},\ \bibinfo {pages}
  {012002} (\bibinfo {year} {2020})},\ \Eprint
  {http://arxiv.org/abs/1909.02845} {arXiv:1909.02845 [hep-ex]} \BibitemShut
  {NoStop}%
\bibitem [{\citenamefont {Sirunyan}\ \emph {et~al.}(2019)\citenamefont
  {Sirunyan} \emph {et~al.}}]{CMS:2018uag}%
  \BibitemOpen
  \bibfield  {author} {\bibinfo {author} {\bibfnamefont {A.~M.}\ \bibnamefont
  {Sirunyan}} \emph {et~al.} (\bibinfo {collaboration} {CMS}),\ }\href
  {\doibase 10.1140/epjc/s10052-019-6909-y} {\bibfield  {journal} {\bibinfo
  {journal} {Eur. Phys. J. C}\ }\textbf {\bibinfo {volume} {79}},\ \bibinfo
  {pages} {421} (\bibinfo {year} {2019})},\ \Eprint
  {http://arxiv.org/abs/1809.10733} {arXiv:1809.10733 [hep-ex]} \BibitemShut
  {NoStop}%
\bibitem [{\citenamefont {Abi}\ \emph {et~al.}(2021)\citenamefont {Abi} \emph
  {et~al.}}]{Muong-2:2021ojo}%
  \BibitemOpen
  \bibfield  {author} {\bibinfo {author} {\bibfnamefont {B.}~\bibnamefont
  {Abi}} \emph {et~al.} (\bibinfo {collaboration} {Muon g-2}),\ }\href
  {\doibase 10.1103/PhysRevLett.126.141801} {\bibfield  {journal} {\bibinfo
  {journal} {Phys. Rev. Lett.}\ }\textbf {\bibinfo {volume} {126}},\ \bibinfo
  {pages} {141801} (\bibinfo {year} {2021})},\ \Eprint
  {http://arxiv.org/abs/2104.03281} {arXiv:2104.03281 [hep-ex]} \BibitemShut
  {NoStop}%
\bibitem [{\citenamefont {Cacciapaglia}\ \emph {et~al.}(2022)\citenamefont
  {Cacciapaglia}, \citenamefont {Cot},\ and\ \citenamefont
  {Sannino}}]{Cacciapaglia:2021gff}%
  \BibitemOpen
  \bibfield  {author} {\bibinfo {author} {\bibfnamefont {G.}~\bibnamefont
  {Cacciapaglia}}, \bibinfo {author} {\bibfnamefont {C.}~\bibnamefont {Cot}}, \
  and\ \bibinfo {author} {\bibfnamefont {F.}~\bibnamefont {Sannino}},\ }\href
  {\doibase 10.1016/j.physletb.2021.136864} {\bibfield  {journal} {\bibinfo
  {journal} {Phys. Lett. B}\ }\textbf {\bibinfo {volume} {825}},\ \bibinfo
  {pages} {136864} (\bibinfo {year} {2022})},\ \Eprint
  {http://arxiv.org/abs/2104.08818} {arXiv:2104.08818 [hep-ph]} \BibitemShut
  {NoStop}%
\bibitem [{\citenamefont {Borsanyi}\ \emph {et~al.}(2020)\citenamefont
  {Borsanyi} \emph {et~al.}}]{Borsanyi:2020mff}%
  \BibitemOpen
  \bibfield  {author} {\bibinfo {author} {\bibfnamefont {S.}~\bibnamefont
  {Borsanyi}} \emph {et~al.},\ }\href {\doibase 10.1038/s41586-021-03418-1}
  {\bibfield  {journal} {\bibinfo  {journal} {Nature}\ } (\bibinfo {year}
  {2020}),\ 10.1038/s41586-021-03418-1},\ \Eprint
  {http://arxiv.org/abs/2002.12347} {arXiv:2002.12347 [hep-lat]} \BibitemShut
  {NoStop}%
\bibitem [{\citenamefont {Aoyama}\ \emph {et~al.}(2020)\citenamefont {Aoyama}
  \emph {et~al.}}]{Aoyama:2020ynm}%
  \BibitemOpen
  \bibfield  {author} {\bibinfo {author} {\bibfnamefont {T.}~\bibnamefont
  {Aoyama}} \emph {et~al.},\ }\href {\doibase 10.1016/j.physrep.2020.07.006}
  {\bibfield  {journal} {\bibinfo  {journal} {Phys. Rept.}\ }\textbf {\bibinfo
  {volume} {887}},\ \bibinfo {pages} {1} (\bibinfo {year} {2020})},\ \Eprint
  {http://arxiv.org/abs/2006.04822} {arXiv:2006.04822 [hep-ph]} \BibitemShut
  {NoStop}%
\bibitem [{\citenamefont {Aaij}\ \emph {et~al.}(2022)\citenamefont {Aaij} \emph
  {et~al.}}]{LHCb:2021trn}%
  \BibitemOpen
  \bibfield  {author} {\bibinfo {author} {\bibfnamefont {R.}~\bibnamefont
  {Aaij}} \emph {et~al.} (\bibinfo {collaboration} {LHCb}),\ }\href {\doibase
  10.1038/s41567-021-01478-8} {\bibfield  {journal} {\bibinfo  {journal}
  {Nature Phys.}\ }\textbf {\bibinfo {volume} {18}},\ \bibinfo {pages} {277}
  (\bibinfo {year} {2022})},\ \Eprint {http://arxiv.org/abs/2103.11769}
  {arXiv:2103.11769 [hep-ex]} \BibitemShut {NoStop}%
\bibitem [{\citenamefont {Sannino}\ \emph {et~al.}(2016)\citenamefont
  {Sannino}, \citenamefont {Strumia}, \citenamefont {Tesi},\ and\ \citenamefont
  {Vigiani}}]{Sannino:2016sfx}%
  \BibitemOpen
  \bibfield  {author} {\bibinfo {author} {\bibfnamefont {F.}~\bibnamefont
  {Sannino}}, \bibinfo {author} {\bibfnamefont {A.}~\bibnamefont {Strumia}},
  \bibinfo {author} {\bibfnamefont {A.}~\bibnamefont {Tesi}}, \ and\ \bibinfo
  {author} {\bibfnamefont {E.}~\bibnamefont {Vigiani}},\ }\href {\doibase
  10.1007/JHEP11(2016)029} {\bibfield  {journal} {\bibinfo  {journal} {JHEP}\
  }\textbf {\bibinfo {volume} {11}},\ \bibinfo {pages} {029} (\bibinfo {year}
  {2016})},\ \Eprint {http://arxiv.org/abs/1607.01659} {arXiv:1607.01659
  [hep-ph]} \BibitemShut {NoStop}%
\bibitem [{\citenamefont {Cacciapaglia}\ \emph {et~al.}(2020)\citenamefont
  {Cacciapaglia}, \citenamefont {Pica},\ and\ \citenamefont
  {Sannino}}]{Cacciapaglia:2020kgq}%
  \BibitemOpen
  \bibfield  {author} {\bibinfo {author} {\bibfnamefont {G.}~\bibnamefont
  {Cacciapaglia}}, \bibinfo {author} {\bibfnamefont {C.}~\bibnamefont {Pica}},
  \ and\ \bibinfo {author} {\bibfnamefont {F.}~\bibnamefont {Sannino}},\ }\href
  {\doibase 10.1016/j.physrep.2020.07.002} {\bibfield  {journal} {\bibinfo
  {journal} {Phys. Rept.}\ }\textbf {\bibinfo {volume} {877}},\ \bibinfo
  {pages} {1} (\bibinfo {year} {2020})},\ \Eprint
  {http://arxiv.org/abs/2002.04914} {arXiv:2002.04914 [hep-ph]} \BibitemShut
  {NoStop}%
\bibitem [{\citenamefont {Yamawaki}\ \emph {et~al.}(1986)\citenamefont
  {Yamawaki}, \citenamefont {Bando},\ and\ \citenamefont
  {Matumoto}}]{Yamawaki:1985zg}%
  \BibitemOpen
  \bibfield  {author} {\bibinfo {author} {\bibfnamefont {K.}~\bibnamefont
  {Yamawaki}}, \bibinfo {author} {\bibfnamefont {M.}~\bibnamefont {Bando}}, \
  and\ \bibinfo {author} {\bibfnamefont {K.-i.}\ \bibnamefont {Matumoto}},\
  }\href {\doibase 10.1103/PhysRevLett.56.1335} {\bibfield  {journal} {\bibinfo
   {journal} {Phys. Rev. Lett.}\ }\textbf {\bibinfo {volume} {56}},\ \bibinfo
  {pages} {1335} (\bibinfo {year} {1986})}\BibitemShut {NoStop}%
\bibitem [{\citenamefont {Sannino}(2016)}]{Sannino:2015yxa}%
  \BibitemOpen
  \bibfield  {author} {\bibinfo {author} {\bibfnamefont {F.}~\bibnamefont
  {Sannino}},\ }\href {\doibase 10.1103/PhysRevD.93.105011} {\bibfield
  {journal} {\bibinfo  {journal} {Phys. Rev. D}\ }\textbf {\bibinfo {volume}
  {93}},\ \bibinfo {pages} {105011} (\bibinfo {year} {2016})},\ \Eprint
  {http://arxiv.org/abs/1508.07413} {arXiv:1508.07413 [hep-ph]} \BibitemShut
  {NoStop}%
\bibitem [{\citenamefont {Hong}\ \emph {et~al.}(2004)\citenamefont {Hong},
  \citenamefont {Hsu},\ and\ \citenamefont {Sannino}}]{Hong:2004td}%
  \BibitemOpen
  \bibfield  {author} {\bibinfo {author} {\bibfnamefont {D.~K.}\ \bibnamefont
  {Hong}}, \bibinfo {author} {\bibfnamefont {S.~D.~H.}\ \bibnamefont {Hsu}}, \
  and\ \bibinfo {author} {\bibfnamefont {F.}~\bibnamefont {Sannino}},\ }\href
  {\doibase 10.1016/j.physletb.2004.07.007} {\bibfield  {journal} {\bibinfo
  {journal} {Phys. Lett. B}\ }\textbf {\bibinfo {volume} {597}},\ \bibinfo
  {pages} {89} (\bibinfo {year} {2004})},\ \Eprint
  {http://arxiv.org/abs/hep-ph/0406200} {arXiv:hep-ph/0406200} \BibitemShut
  {NoStop}%
\bibitem [{\citenamefont {Kaplan}\ and\ \citenamefont
  {Georgi}(1984)}]{Kaplan:1983fs}%
  \BibitemOpen
  \bibfield  {author} {\bibinfo {author} {\bibfnamefont {D.~B.}\ \bibnamefont
  {Kaplan}}\ and\ \bibinfo {author} {\bibfnamefont {H.}~\bibnamefont
  {Georgi}},\ }\href {\doibase 10.1016/0370-2693(84)91177-8} {\bibfield
  {journal} {\bibinfo  {journal} {Phys. Lett. B}\ }\textbf {\bibinfo {volume}
  {136}},\ \bibinfo {pages} {183} (\bibinfo {year} {1984})}\BibitemShut
  {NoStop}%
\bibitem [{\citenamefont {Contino}\ \emph {et~al.}(2003)\citenamefont
  {Contino}, \citenamefont {Nomura},\ and\ \citenamefont
  {Pomarol}}]{Contino:2003ve}%
  \BibitemOpen
  \bibfield  {author} {\bibinfo {author} {\bibfnamefont {R.}~\bibnamefont
  {Contino}}, \bibinfo {author} {\bibfnamefont {Y.}~\bibnamefont {Nomura}}, \
  and\ \bibinfo {author} {\bibfnamefont {A.}~\bibnamefont {Pomarol}},\ }\href
  {\doibase 10.1016/j.nuclphysb.2003.08.027} {\bibfield  {journal} {\bibinfo
  {journal} {Nucl. Phys. B}\ }\textbf {\bibinfo {volume} {671}},\ \bibinfo
  {pages} {148} (\bibinfo {year} {2003})},\ \Eprint
  {http://arxiv.org/abs/hep-ph/0306259} {arXiv:hep-ph/0306259} \BibitemShut
  {NoStop}%
\bibitem [{\citenamefont {Weinberg}(1976)}]{Weinberg:1975gm}%
  \BibitemOpen
  \bibfield  {author} {\bibinfo {author} {\bibfnamefont {S.}~\bibnamefont
  {Weinberg}},\ }\href {\doibase 10.1103/PhysRevD.19.1277} {\bibfield
  {journal} {\bibinfo  {journal} {Phys. Rev. D}\ }\textbf {\bibinfo {volume}
  {13}},\ \bibinfo {pages} {974} (\bibinfo {year} {1976})},\ \bibinfo {note}
  {[Addendum: Phys.Rev.D 19, 1277--1280 (1979)]}\BibitemShut {NoStop}%
\bibitem [{\citenamefont {Susskind}(1979)}]{Susskind:1978ms}%
  \BibitemOpen
  \bibfield  {author} {\bibinfo {author} {\bibfnamefont {L.}~\bibnamefont
  {Susskind}},\ }\href {\doibase 10.1103/PhysRevD.20.2619} {\bibfield
  {journal} {\bibinfo  {journal} {Phys. Rev. D}\ }\textbf {\bibinfo {volume}
  {20}},\ \bibinfo {pages} {2619} (\bibinfo {year} {1979})}\BibitemShut
  {NoStop}%
\bibitem [{\citenamefont {Sannino}(2009)}]{Sannino:2009za}%
  \BibitemOpen
  \bibfield  {author} {\bibinfo {author} {\bibfnamefont {F.}~\bibnamefont
  {Sannino}},\ }\href@noop {} {\bibfield  {journal} {\bibinfo  {journal} {Acta
  Phys. Polon. B}\ }\textbf {\bibinfo {volume} {40}},\ \bibinfo {pages} {3533}
  (\bibinfo {year} {2009})},\ \Eprint {http://arxiv.org/abs/0911.0931}
  {arXiv:0911.0931 [hep-ph]} \BibitemShut {NoStop}%
\bibitem [{\citenamefont {Dugan}\ \emph {et~al.}(1985)\citenamefont {Dugan},
  \citenamefont {Georgi},\ and\ \citenamefont {Kaplan}}]{Dugan:1984hq}%
  \BibitemOpen
  \bibfield  {author} {\bibinfo {author} {\bibfnamefont {M.~J.}\ \bibnamefont
  {Dugan}}, \bibinfo {author} {\bibfnamefont {H.}~\bibnamefont {Georgi}}, \
  and\ \bibinfo {author} {\bibfnamefont {D.~B.}\ \bibnamefont {Kaplan}},\
  }\href {\doibase 10.1016/0550-3213(85)90221-4} {\bibfield  {journal}
  {\bibinfo  {journal} {Nucl. Phys. B}\ }\textbf {\bibinfo {volume} {254}},\
  \bibinfo {pages} {299} (\bibinfo {year} {1985})}\BibitemShut {NoStop}%
\bibitem [{\citenamefont {Agashe}\ \emph {et~al.}(2005)\citenamefont {Agashe},
  \citenamefont {Contino},\ and\ \citenamefont {Pomarol}}]{Agashe:2004rs}%
  \BibitemOpen
  \bibfield  {author} {\bibinfo {author} {\bibfnamefont {K.}~\bibnamefont
  {Agashe}}, \bibinfo {author} {\bibfnamefont {R.}~\bibnamefont {Contino}}, \
  and\ \bibinfo {author} {\bibfnamefont {A.}~\bibnamefont {Pomarol}},\ }\href
  {\doibase 10.1016/j.nuclphysb.2005.04.035} {\bibfield  {journal} {\bibinfo
  {journal} {Nucl. Phys. B}\ }\textbf {\bibinfo {volume} {719}},\ \bibinfo
  {pages} {165} (\bibinfo {year} {2005})},\ \Eprint
  {http://arxiv.org/abs/hep-ph/0412089} {arXiv:hep-ph/0412089} \BibitemShut
  {NoStop}%
\bibitem [{\citenamefont {Liu}\ \emph {et~al.}(2018)\citenamefont {Liu},
  \citenamefont {Low},\ and\ \citenamefont {Yin}}]{Liu:2018vel}%
  \BibitemOpen
  \bibfield  {author} {\bibinfo {author} {\bibfnamefont {D.}~\bibnamefont
  {Liu}}, \bibinfo {author} {\bibfnamefont {I.}~\bibnamefont {Low}}, \ and\
  \bibinfo {author} {\bibfnamefont {Z.}~\bibnamefont {Yin}},\ }\href {\doibase
  10.1103/PhysRevLett.121.261802} {\bibfield  {journal} {\bibinfo  {journal}
  {Phys. Rev. Lett.}\ }\textbf {\bibinfo {volume} {121}},\ \bibinfo {pages}
  {261802} (\bibinfo {year} {2018})},\ \Eprint
  {http://arxiv.org/abs/1805.00489} {arXiv:1805.00489 [hep-ph]} \BibitemShut
  {NoStop}%
\bibitem [{\citenamefont {Hashimoto}(2017)}]{Hashimoto:2017jvc}%
  \BibitemOpen
  \bibfield  {author} {\bibinfo {author} {\bibfnamefont {M.}~\bibnamefont
  {Hashimoto}},\ }\href {\doibase 10.1103/PhysRevD.96.035020} {\bibfield
  {journal} {\bibinfo  {journal} {Phys. Rev. D}\ }\textbf {\bibinfo {volume}
  {96}},\ \bibinfo {pages} {035020} (\bibinfo {year} {2017})},\ \Eprint
  {http://arxiv.org/abs/1704.02615} {arXiv:1704.02615 [hep-ph]} \BibitemShut
  {NoStop}%
\bibitem [{\citenamefont {Cepeda}\ \emph {et~al.}(2019)\citenamefont {Cepeda}
  \emph {et~al.}}]{Cepeda:2019klc}%
  \BibitemOpen
  \bibfield  {author} {\bibinfo {author} {\bibfnamefont {M.}~\bibnamefont
  {Cepeda}} \emph {et~al.},\ }\href {\doibase 10.23731/CYRM-2019-007.221}
  {\bibfield  {journal} {\bibinfo  {journal} {CERN Yellow Rep. Monogr.}\
  }\textbf {\bibinfo {volume} {7}},\ \bibinfo {pages} {221} (\bibinfo {year}
  {2019})},\ \Eprint {http://arxiv.org/abs/1902.00134} {arXiv:1902.00134
  [hep-ph]} \BibitemShut {NoStop}%
\bibitem [{\citenamefont {Abada}\ \emph {et~al.}(2019)\citenamefont {Abada}
  \emph {et~al.}}]{FCC:2018byv}%
  \BibitemOpen
  \bibfield  {author} {\bibinfo {author} {\bibfnamefont {A.}~\bibnamefont
  {Abada}} \emph {et~al.} (\bibinfo {collaboration} {FCC}),\ }\href {\doibase
  10.1140/epjc/s10052-019-6904-3} {\bibfield  {journal} {\bibinfo  {journal}
  {Eur. Phys. J. C}\ }\textbf {\bibinfo {volume} {79}},\ \bibinfo {pages} {474}
  (\bibinfo {year} {2019})}\BibitemShut {NoStop}%
\end{thebibliography}%


\end{document}